\documentclass[prl,twocolumn,preprintnumbers,superscriptaddress,amsmath,amssymb]{revtex4}

\usepackage{graphicx}
\usepackage{color}
\usepackage{dcolumn}
\usepackage{tabularx}
\usepackage[french]{babel}

\begin{document}

\title{Single photon emission from graphene quantum dots at room temperature}

\author{S. Zhao}
\affiliation{Laboratoire Aim\'e Cotton, CNRS, Univ. Paris-Sud, ENS Cachan, Universit\'e Paris Saclay bat 505 campus d'Orsay, 91405 Orsay cedex France}
\author{J. Lavie}
\affiliation{LICSEN, NIMBE, CEA, CNRS, Universit\'e Paris-Saclay}
\author{L. Rondin}
\affiliation{Laboratoire Aim\'e Cotton, CNRS, Univ. Paris-Sud, ENS Cachan, Universit\'e Paris Saclay bat 505 campus d'Orsay, 91405 Orsay cedex France}

\author{L. Orcin-Chaix}
\affiliation{Laboratoire Aim\'e Cotton, CNRS, Univ. Paris-Sud, ENS Cachan, Universit\'e Paris Saclay bat 505 campus d'Orsay, 91405 Orsay cedex France}
\affiliation{LICSEN, NIMBE, CEA, CNRS, Universit\'e Paris-Saclay}
\author{C. Diederichs}
\author{Ph. Roussignol}
\author{Y. Chassagneux}
\author{C. Voisin}
\affiliation{
Laboratoire Pierre Aigrain, D\'epartement de physique de l'ENS, \'Ecole normale sup\'erieure, PSL Research University, Universit\'e Paris Diderot, Sorbonne Paris Cit\'e, Sorbonne Universit\'es, UPMC Univ. Paris 06, CNRS, 75005 Paris, France}
\author{K. M\"{u}llen}
\author{A. Narita}
\affiliation{Max Planck Institute for Polymer Research, Ackermannweg 10, 55128 Mainz, Germany}
\author{S. Campidelli}
\affiliation{LICSEN, NIMBE, CEA, CNRS, Universit\'e Paris-Saclay}
\author{J.S. Lauret}
\email{lauret@ens-paris-saclay.fr}
\affiliation{Laboratoire Aim\'e Cotton, CNRS, Univ. Paris-Sud, ENS Cachan, Universit\'e Paris Saclay bat 505 campus d'Orsay, 91405 Orsay cedex France}

\maketitle

\renewcommand\figurename{\textbf{Figure}}

\textbf{In the field of condensed matter, graphene plays a central role as an emerging material for nanoelectronics. Nevertheless, graphene is a semimetal, which constitutes a severe limitation for some future applications. Therefore, a lot of efforts are being made to develop semiconductor materials whose structure is compatible with the graphene lattice. In this perspective, little pieces of graphene represent a promising alternative~\cite{Mullen2014a, Wu2007}. In particular, their electronic, optical and spin properties can be in principle controlled by designing their size, shape and edges~\cite{Tomovic2004, Debije2004, Yan2010, Konishi2010}. As an example, graphene nanoribbons with zigzag edges have localized spin polarized states~\cite{Son2006, Ruffieux2016}. Likewise, singlet-triplet energy splitting can be chosen by designing the structure of graphene quantum dots~\cite{Li2015}. Moreover, bottom-up molecular synthesis put these potentialities at our fingertips~\cite{Tomovic2004, Yan2010}. Here, we report on a single emitter study that directly addresses the intrinsic properties of a single graphene quantum dot. In particular, we show that graphene quantum dots emit single photons at room temperature with a high purity, a high brightness and a good photostability. These results pave the way to the development of new quantum systems based on these nanoscale pieces of graphene.}

\begin{figure*}
\includegraphics[scale=0.195]{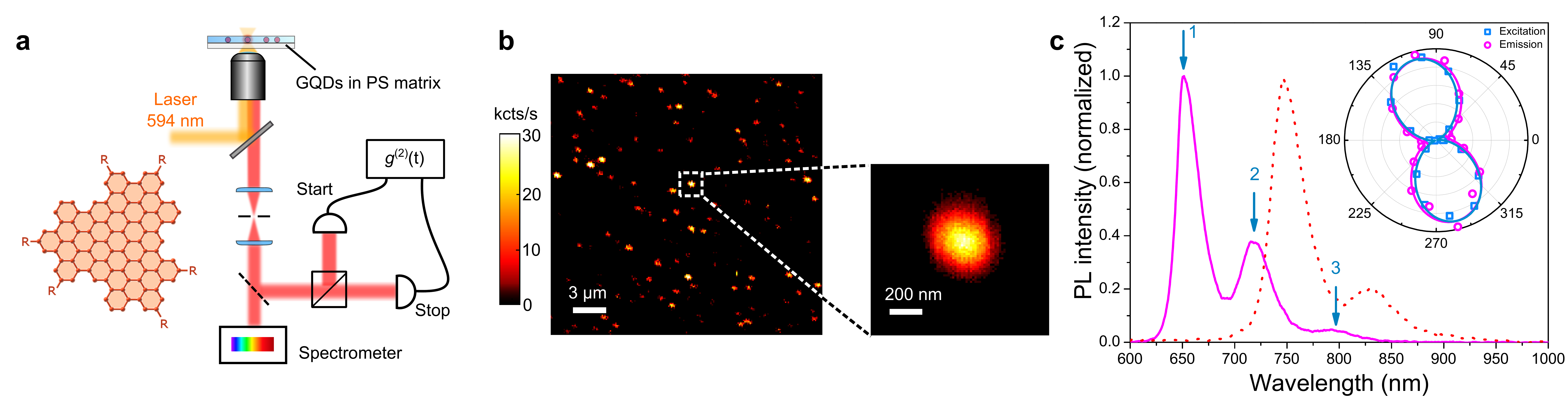}
\caption{\textbf{Photoluminescence of single GQDs.} \textbf{a}, Chemical structure of the C$_{96}$ GQD. $R$ stands for C$_{12}$H$_{25}$. Scheme of the microphotoluminescence setup. \textbf{b}, 20$\times$20~$\mu$m$^2$ PL map of GQDs in polystyrene matrix. The color bar represents the number of counts per second on the APD. The zoom shows a diffraction limited spot that can be fitted with a 2D Gaussian function leading to a $1/e^{2}$ diameter of $\sim$600~nm. \textbf{c}, Room temperature PL spectra of a single C$_{96}$ GQD (solid line) and of a single C$_{96}$Cl GQD (dotted line). Inset: polarization diagram in excitation (blue) and emission (red).}
\label{fig1}
\end{figure*}

To date, graphene quantum dots (GQDs) have been mostly synthesized by "top-down" approaches, such as the oxidation of carbon fibers or graphene ~\cite{Li2013a,Bacon2014}. These methods present important drawbacks that make them not suitable to produce emitters with well defined properties. In particular, they do not allow to control neither the size nor the chemical nature of the edges of the dots. Therefore, their optical properties are dominated by defect states instead of the intrinsic quantum confined states of the nanographene~\cite{Xu2013}. Fortunately, over the last two decades, "bottom-up" synthesis have been developed opening the way to a precise control of the GQD structure~\cite{Wu2007}. Several sizes and shapes of GQDs have been already synthesized \cite {Mullen2014a}. However, to date these materials often face problems that hinder the possibility to make the link between their structure and their intrinsic properties, especially in solid state or at the single-emitter level~\cite{Mullen2014a, Zhao2017}. In this paper, we study the photoluminescence of  single GQDs synthesized by this bottom-up approach. We demonstrate the access to their intrinsic properties. Notably, we highlight that all the measured GQDs emit single photons at room temperature with a high purity, a high brightness and a good photostability putting them in the foreground of the research on new single emitters.

The chemical structure of the GQD studied here is displayed on Fig.~\ref{fig1}a. It is made of 96 sp$^2$ carbon atoms arranged in a triangular shape which leads to lateral sizes of $\sim$2~nm. Six alkyl chains (R=C$_{12}$H$_{25}$) were introduced at the edges to enhance the solubility of the GQDs. The details of the synthesis have been already reported~\cite{Tomovic2004}. Briefly, the C$_{96}$ GQD is synthesized in two steps from the 1,3,5-triethynylbenzene and 2,5-diphenyl-3,4-di(4-dodecylphenyl)-cyclopentadien-1-one via Diels-Alder cycloaddition followed by oxidative cyclodehydrogenation in the presence of FeCl$_{3}$ (See Extended Data Fig.~\ref{EAFigChimie1}). The intermediate compounds are fully characterized and the GQD is characterized by MALDI-TOF spectrometry to confirm its complete dehydrogenation (See Extended Data Fig.~\ref{EAFigChimie2}). The GQDs are dispersed in 1,2,4-trichlorobenzene and then mixed with a  solution of polystyrene (PS). The mixture is subsequently spin-coated on a coverslip to perform the optical experiments, using the experimental setup described in Fig.~\ref{fig1}a. Complementary details about the sample preparation can be found in the Methods section.

Fig.~\ref{fig1}b shows an example of a photoluminescence (PL) map of GQDs embedded in PS matrix for the highest dilution used in this study. One can observe bright spots (30~kcounts/s with a 200~nW excitation) with diffraction limited size. The PL spectrum acquired on such a spot is displayed on Fig.~\ref{fig1}c. For this particular GQD, the spectrum is composed of three lines, noted 1, 2 and 3, centered at 653~nm, 719~nm and 797~nm, respectively. The full width at half maximum of the main line is of the order of 27~nm (80~meV). Note that the wavelength as well as the relative intensity of the PL lines slightly vary from one GQD to another, which is certainly due to differences in their local environments. Also, averaging over 25 GQDs gives an energy splitting of $170\pm5$~meV between lines 1 and 2, and $165\pm10$~meV between lines 2 and 3. Moreover, as mentioned in the introduction, one of the great potential of GQDs lies in the precise tuning of their electronic properties through the control of their structure. As a first example, the dotted curve in Fig~\ref{fig1}c shows the PL spectrum of a single C$_{96}$ GQD whose emission wavelength has been tuned by precise chemical functionalization of the edges with chlorine atoms (see Extended Data Fig.\ref{EAFigC96Cl} for the chemical structure). Here the functionalization leads to an almost 100~nm rigid redshift of the main PL line. This observation is in agreement with theoretical predictions that show a decrease of the optical gap of C$_{96}$ GQD due to the electronegativity of the chlorine atoms~\cite{Tan2013}. This result illustrates how the intrinsic properties of GQDs can be controlled by the synthesis, and opens the way to even finer tunning.

In the following, we focus on C$_{96}$ GQDs with the C$_{12}$H$_{25}$ alkyl chains. The exact nature of the quantum states at the origin of the three PL lines is still an open question that needs a specific study. To comment on the likely mechanisms that could be responsible for these spectra, we performed photoluminescence excitation (PLE) experiments in solution (See Extended Data Fig.~\ref{EAFig1}). The PLE curves detected on the two highest energy PL lines of GQD superimpose well with the absorption spectrum of the solution.  Besides, the PLE highlights the existence of two states at 580~nm and 630~nm with small oscillator strengths in comparison with the main absorption line centered at 475~nm. This observation is in good agreement with calculated absorption spectra. Indeed, theoretical works on this family of objects show an intense line at high energy called $\beta$ band in the Clar's notation, accompanied by lower lying dark states, the so-called $\alpha$ and $p$ bands~\cite{Rieger2010, Cocchi2014}. The coupling of the $\alpha$ and $p$ bands with the vibrations of the lattice can then lead to a brightening of these low energy states~\cite{Beljonne}. Therefore, the two PLE bands at 580~nm and 630~nm can tentatively be attributed to the intrinsic $\alpha$ and $p$ bands while the main one at 475~nm would be assigned to the $\beta$ band. However, both PLE and PL spectra are more complex than this first description. Especially, the PL spectrum of single GQDs is composed of several lines. The almost 170~meV energy splitting between each PL line suggests potential vibronic effects. It can be seen here that further studies, including calculations, are necessary for a full understanding of the electronic structure. An interesting point will be for instance to search for signatures of the band diagram of graphene, such as valley structure. Finally, the polarization response of a single GQD  is shown on the inset of Fig.~\ref{fig1}c. Here, the emission polarization profile is recorded on the line~1 at $\sim$650~nm. It is linearly polarized at a fixed direction certainly related to the geometry of the GQD. Likewise the excitation diagram is also linearly polarized in the same direction than the emission one. Therefore, it can be concluded that absorption and main emission dipoles are parallel. Further investigations, including theoretical modeling, are also needed to explain this experimental observation.

\begin{figure*}
\includegraphics[scale=0.9]{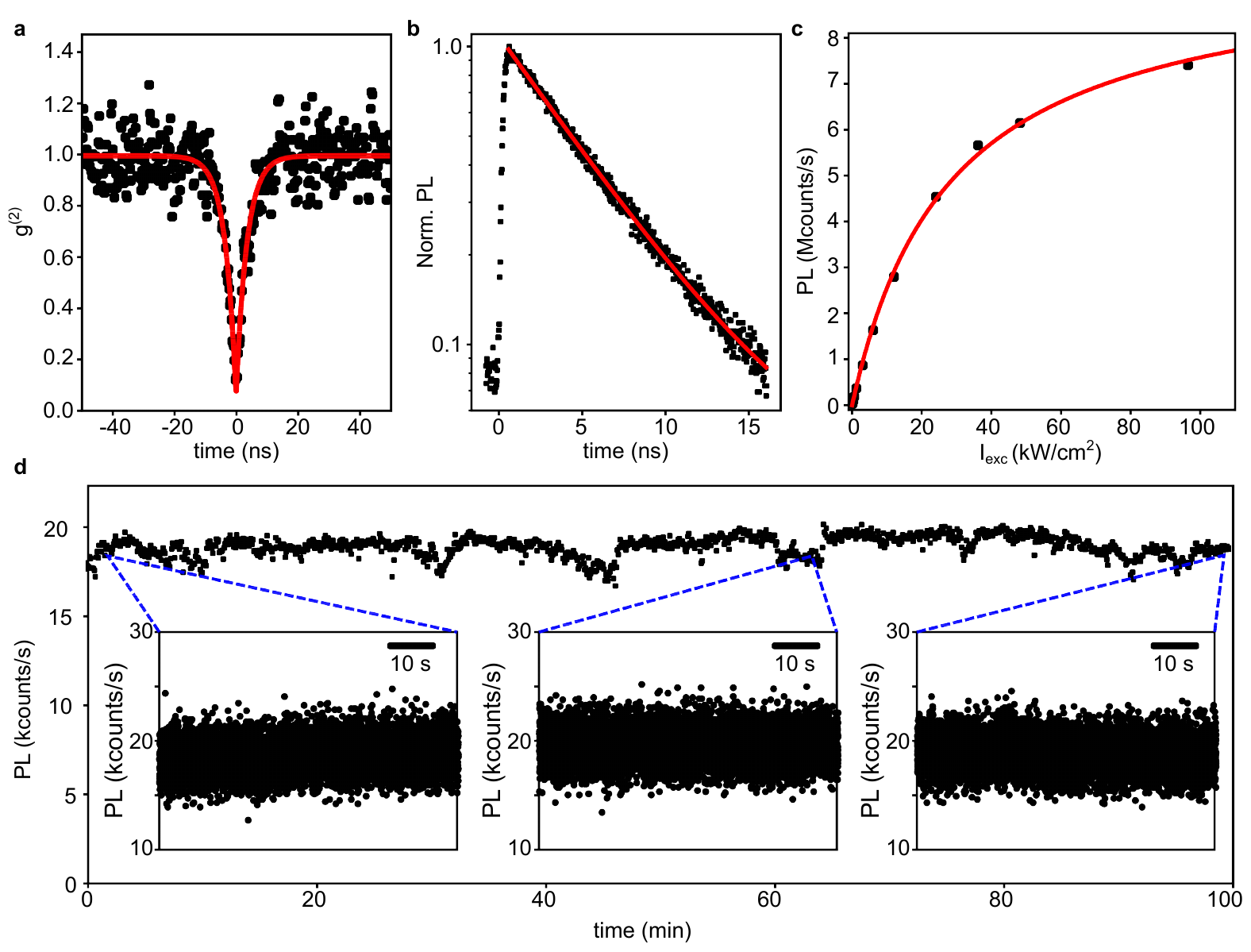}
\caption{\textbf{Photophysics of a single GQD. a}, Second order correlation function $g^2(\tau)$ recorded from a diffraction limited spot such as the one of Fig.~\ref{fig1} (black dots), showing a strong antibunching. A fit (red line) with a function $1-(1-b)e^{-|\tau|/\tau _{1}}$ yields $g^2(0)=0.05 \pm 0.05$ and a characteristic time of $\tau_1\sim 3.5$~ns (FWHM of the IRF of the detector $\sim0.9$~ns). \textbf{b}, Time-resolved PL of a single GQD (black dots) detected on the whole spectrum, fitted by a monoexponential decay (red line) with a time constant $\tau=5.37$~ns.  \textbf{c}, Saturation curve of a single GQD (black dots) as a function of the pump power. A fit by eq.~(\ref{eq:Sat}) (red line) leads to a saturation power density of 28~kW$\cdot$cm$^{-2}$ and a saturation intensity $I_\mathrm{sat}\sim 9.7$~Mcounts/s. \textbf{d}, PL time trace of a stable GQD over 100~minutes with a binning time of 200~ms. Fluctuations are due to setup instabilities. Zooms are shown on shorter timescale with a binning time of 10~ms.}
\label{fig2}
\end{figure*}

In order to identify the number of emitters associated with such diffraction limited spot and spectrum, we measured the second order correlation function ($g^{(2)}(\tau)$) at room temperature on photons from the entire spectrum. As displayed in Fig.~\ref{fig2}a,  the strong antibunching observed at zero delay, $g^{(2)}(0)<0.5$, is a  proof that a single emitter is detected. This is in strong contrast with the results of such experiments performed on "top-down" GQDs where no antibunching is observed~\cite{Xu2013}. In these earlier studies, the absence of antibunching is interpreted as a consequence of the extrinsic nature of the states at the origin of the luminescence~\cite{Xu2013}. We have performed measurements on 30 GQDs, all of them leading to $g^{(2)}(0)<0.1$ (see examples on Extended Data Fig.~\ref{EAFig6}). Moreover, these correlation measurements being performed by integrating all the wavelengths on the detector, it implies that the spectrum described above indeed arises from a single GQDs and not from several objects. Moreover, the weak value observed for the $g^{(2)}(0)$ is an indication of the good purity of single photons emission associated with single GQD. This result enforces GQD as an interesting alternative to other single emitters in 2D materials, such as defects in WSe$_2$~\cite{Koperski2015, Srivastava2015, Chakraborty2015, He2015} or in h-BN~\cite{Li2017, Martinez2016c, Tran2016}.

The assurance that we are observing single objects allows us to characterise other important properties. First, another figure of merit of a single quantum emitter is its brightness. A saturation curve of a single GQD is displayed on Fig.~\ref{fig2}c. The intensity is fitted by
\begin{equation}
    R= R_\mathrm{sat}/(1+\frac{I_\mathrm{sat}}{I_\mathrm{exc}})\, ,
    \label{eq:Sat}
\end{equation}
with $R_\mathrm{sat}$ the count rate at saturation, $I_\mathrm{sat}$ the incident power at saturation and $I_\mathrm{exc}$ the incident power. The fit leads to $R_\mathrm{sat}\sim  9.7$~Mcounts/s,  and $I_\mathrm{sat} = 28$~kW$\cdot$cm$^{-2}$. This value of count rate at saturation has to be compared to other quantum emitters. For instance,  the wide use of single NV center in bulk diamond, with a well-documented count rate, make it a good point of comparison for new quantum emitters~\cite{Martinez2016c}. In the present case, we measured a $R_\mathrm{sat} \approx 0.3$~Mcounts/s for a single NV center in a 111-diamond (see Extended Data Fig.~\ref{EAFig2}), in good agreement with the literature~\cite{Lesik2014}. The intensity at saturation of GQDs is thus $\sim30$ times higher. Therefore, it puts GQDs in the highest values of brightness among other quantum emitters \cite{Tran2016, Martinez2016c}. Secondly, the question of the photostability of the quantum emitter is also an important issue. In this perspective, GQDs have also good properties. Indeed, photostability up to hours have been observed for an incident power of 200~nW (0.12~kW.cm$^{-2}$) and an emission rate of $\sim$20~kcounts/s (see Fig.~\ref{fig2}d). This is very encouraging since no particular care has been taken for the preparation of the sample. Moreover, Fig.~\ref{fig2}d shows that no blinking is occurring on the time trace of the luminescence of this GQD, the histogram of intensity fitting well to a normal distribution (see Extended Data Fig.~\ref{EAFig3}). This observation is in strong contrast with numerous single emitters that undergo quantum jumps between different states due, for instance, to quantum confined stark effect~\cite{Empedocles1997}. Nevertheless, the luminescence of GQDs end up disappearing. Extended Data Fig.~\ref{EAFig4} shows two different time traces around the time of bleaching. In the first one the luminescence drops down to zero sharply. On the contrary, the second one goes through an intermediate "grey" state before ending up to zero. These two behaviors are representative of what we observed on all the GQDs. The discrete intensity jumps are also characteristic of single emitter experiments. Further work is needed to understand the mechanisms behind such behaviors. In particular, it would be interesting to investigate whether the GQD emission is irreversibly quenched  or if they are able to reemit light after a certain time. This could lead to a better control of the local environment of GQDs in the view of optimizing their long term photostability. Furthermore, we have measured the luminescence lifetime on a single GQD. Fig~\ref{fig2}b shows the time-resolved photoluminescence (TR-PL) curve for a single GQD. Here, the signal can be fitted by a mono-exponential decay with a time constant $\tau\sim5.37$~ns. We have performed such experiments on several GQDs. The TR-PL signal is always mono-exponential with relaxation times ranging from 3 to 5.5~ns.

\begin{figure}
\includegraphics[scale=1.075]{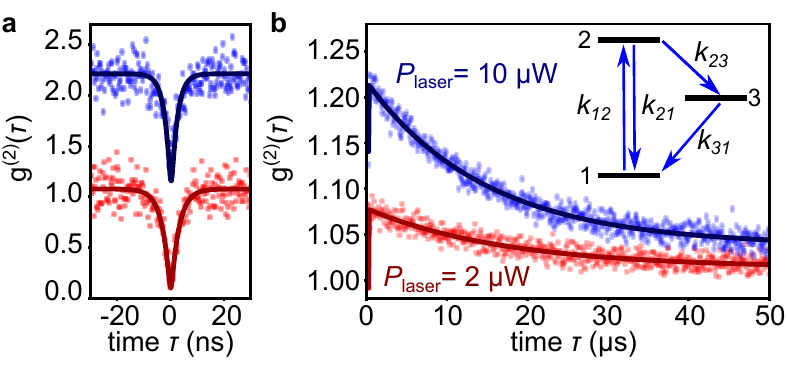}
\caption{\textbf{Photons bunching of a single GQD.} \textbf{a-b} $g^{(2)}$ functions
of a single GQD, for two different excitation powers, 2~$\mu$W (red square) and
10~$\mu$W (blue diamond).  The curves have been shifted for clarity.\textbf{a} Zoom on short delays; \textbf{b} Full timescale observation. A fit (solid line) by the function $1-(1+a)e^{-\lambda_1 |\tau|}+ae^{-\lambda_2 |\tau|}$, convolved by the time response of the detector allows extracting the relevant photophysical parameters. The three level system used as a model is shown in the inset.}
\label{fig3}
\end{figure}

Finally, as a first step toward probing the potential of GQDs for spin physics~\cite{Li2010}, and in analogy with observations in single molecules experiments, we chose to study the $g^{(2)}$ function on a longer timescale. Indeed, such measurement allows highlighting intersystem crossing (ISC) dynamics between singlet and triplet states~\cite{Bernard1993}. This approach is also supported by calculations showing that at least one triplet state is lying few hundreds of meV below the singlet state ~\cite{Li2015, Schumacher2011}. Finally, this experiment is particularly suitable to extract details about GQDs' photophysics. The $g^{(2)}$ functions recorded under different excitation density both at short and long time delays are shown respectively in Fig.~\ref{fig3}a and ~\ref{fig3}b. At short delays, one observes a reduction of the dynamics of the $g^{(2)}$ function when the pump is increased. Likewise, at longer delays photon bunching ($g^{(2)} > 1$) is observed with a relaxation down to $g^{(2)}=1$ within tens of microseconds. Assuming a three-level system model, as shown in the inset of Fig. 3b, we can fit these data set to extract relevant photophysical parameters. In particular, the relaxation rates $k_{21}\sim 0.28$~ns$^{-1}$ and $k_{31}\sim0.057$~$\mu$s$^{-1}$ have been extracted (see Methods section for the details of the photophysics). Note that the latter characteristics rate ($k_{31}$) is of the same order of what is typically observed for small molecules in the context of ISC. Therefore, it is reasonable to attribute this metastable state to a triplet state of C$_{96}$. Finally, the modelization leads also to an estimation of the absorption cross section of the C$_{96}$ GQD ($\sigma\simeq 1.0\times 10^{-14}$~cm$^{2}$) as well as of the PL quantum efficiency $\eta_Q$ above $35\%$ (see Methods section).

To conclude, the outstanding properties of GQDs: single photon emission, brightness and photostability show that they are very promising materials for applications requiring single emitters. More generally, the results reported here demonstrate that the high potential of GQDs revealed by theory is now accessible experimentally. In particular, the fact that the emission of a single GQD is observable and tunable pave the way to studies linking their properties to their structure. These results show that the tuning of the properties through the control of the structure is indeed achievable. Therefore, one can now really imagine to perform engineering of other properties such as the spin structure in order to rend it optically detectable and controllable.

\bibliographystyle{naturemag}
\bibliography{./BiblioGQD_v5}
\vspace{0.5cm}
\textbf{Acknowledgements} The authors are grateful to D. Beljonne, J.-F. Roch and V. Jacques for discussions. The authors thank C. Allain for her help on the PLE experiments. This work was partly funded by the JST-ANR program TMOL "Molecular Technology" project MECANO (ANR-14-JTIC-0002-01) and the ANR projet MAGMA (ANR-16-CE29-0027-01) and by a public grant overseen by the French National Research Agency (ANR) as part of the "Investissements d'Avenir" program (Labex NanoSaclay, reference: ANR-10-LABX-0035) and by the GDR-I GNT. J.S.L is partially funded by "Institut Universitaire de France".\\

\textbf{Authors contributions}\\
SZ and LR have performed the optical experiments. JL and LOC have performed the synthesis of GQDs.  SC have supervised the synthesis. LR and JSL have supervised the optical experiments and the analysis of the data. All the authors have discussed on the results and participated to the writing of the paper. \\

\textbf{Methods}\\

\textbf{Techniques}. MALDI-OF spectra were performed with a Perspective Biosystems Voyager DE-STR at the I.C.S.N. (CNRS of Gif-sur-Yvette). 1H-NMR spectra were recorded on Bruker AC-300 spectrometer (300MHz) with CDCl3 as reference solvent.\\

\textbf{Materials.} Chemicals were purchased from Aldrich and were used as received. The 3,4-bis(4-dodecylphenyl)-2,5-diphenylcyclopentadienone was prepared according to the method described in the literature~\cite{Wehmeier2001}. Solvents were purchased from Aldrich, VWR or ThermoFisher and were used as received.\\

\textbf{Synthesis.} Dendrimer \textbf{3}. 3,4-bis(4-dodecylphenyl)-2,5-diphenylcyclopentadienone (450 mg, 0.62 mmol) and the 1,3,5-triethynylbenzene (25 mg, 0.17 mmol) were introduced in a dry Schlenck flask. After two vacuum-argon cycles, o-xylene (2~mL) were added and the reaction was left at 180\degres C  overnight. The suspension was diluted in dichloromethane (2~mL) and the solution was precipitated in cold ethanol (200~mL). After filtration on PTFE membrane (0.2 $\mu$m), the product was redispersed in hot methanol to solublize the excess of cyclopentadienone, filtered on PTFE membrane and washed with methanol. Finally, 220~mg of yellowish powder were obtained (57\% yield). 1H NMR (300MHz, CDCl3): 7.10-6.60 (60H, m, Ar-H); 2.37 (12H, q, J=5.6 Hz, 6CH$_2$(Ar)dodecyl); 1.27 (120H, s, 60CH$_2$-dodecyl); 0.90 (18H, t, J=5.6 Hz, 6CH$_3$-dodecyl). MS (MALDI-TOF): m/z (\%) calcd for C$_{168}$H$_{210}$: 2227.64; found: 2227.60 (M$^{+}$).\\
\textbf{C$_{96}$} quantum dot. Dendrimer \textbf{3} (50~mg, 22.4~$\mu$mol) was dispersed in non-stabilized dichloromethane (40~mL) in a two-necked round bottom flask of 100~mL. Separetly, FeCl$_3$ (458~mg, 2.8 mmol) was dissolved in anhydrous nitromethane in a gloves box (2~mL) and then added to the solution of dendrimer. The solution was left 18~hours under argon coming from a two necked round bottom flask filled with dichloromethane in which argon was bubbling. The evolution of the reaction was followed with MALDI-TOF mass spectroscopmetry after quenching of an aliquot with methanol. When the reaction was finished, the solution was quenched with methanol (40~mL) and then filtered on PTFE membrane (0.2~$\mu$m). 46~mg of a black powder were obtained as the pure product (95\% yield). MS (MALDI-TOF): m/z (\%) calcd for C$_{168}$H$_{174}$: 2191.36; found: 2191.87 (M$^{+}$).\\

\textbf{Sample preparation.} GQDs powder was dispersed in 1,2,4-trichlorobenzene (TCB) by stirring at least 24h at room temperature. For single-molecule measurements,  a 2~mL diluted GQDs solution (0.001~mg.mL$^{-1}$) was mixed with a 2 mL 1,2,4-trichlorobenzene solution (0.08 mg/mL) of polystyrene (PS). Approximately 20~$\mu$L of the GQD-PS-TCB solution was spin-coated on an oxygen-plasma-treated glass coverslip for 180~s at 2000~r.p.m. The sample was dried by heating to 90~$^0$C for 1~h on a hot plate.\\

\textbf{Optical measurements.}
Optical experiments were performed on a home-built micro-PL setup under ambient conditions, as shown in Fig.~\ref{fig1}a. The excitation source was a continuous-wave diode laser at 594~nm (Cobolt, Mambo 100) with linear polarization. The excitation laser was focused onto the sample with a high numerical aperture oil-immersion microscope objective (NA~=~1.42, Olympus PLAPON 60XO) mounted on a piezoelectric XYZ scanner (Mad City Labs Inc.). The luminescence light was collected by the same objective and filtered from the residual excitation laser using a dichroic mirror (zt 594 RDC, Chroma) and a long-pass filter (FELH0600, Thorlabs). The collected luminescence was then focused on a 50-$\mu$m-diameter pinhole and finally directed either into a spectrometer (SP-2358, Princeton Instruments) coupled with a cooled CCD camera (PyLoN:100BR eXcelon, Princeton Instruments) or into two silicon avalanche photodiodes (SPCM-AQR-13, PerkinElmer) mounted in a Hanbury Brown and Twiss (HBT) configuration. Short-time-scale second-order photon correlation measurements were done using a time-correlated single photon counting module (PicoHarp300, PicoQuant). Long-time-scale second-order photon correlation measurements were done using a wide-range time digitizer (P7887, FastComtec). For time-resolved PL measurements, the sample was excited using a supercontinuum laser (Fianium) tuned at 580~nm by an acousto-optic tunable filter systems with a $~$6 ps pulse width and a 60~MHz repetition rate.\\

\textbf{Three-level system.} We now model the GQD response within the framework of a three-level system: ground state (level 1), excited state (level 2) and metastable state (level 3) as shown in Fig.~\ref{fig3} of the main text. The transition rate from level $n$ to level $m$ is given by $k_{nm}$ (with $n,m = 1,2,3$). Assuming that  $k_{23}$ and $k_{31}$ are very small compared to $k_{21}$, the normalized second-order auto-correlation function $g^{(2)}(t)$ can be expressed as \cite{Novotny2012}:
\begin{center}
$g^{(2)}(t)=1-(1+a)e^{-\lambda _{1}t}+ae^{-\lambda _{2}t}$~~~~(1)

\end{center}
with following expressions for the parameters $a, \lambda _{1},$ and $\lambda _{2}$:
\begin{center}
$\lambda _{1}=k_{12}+k_{21}$\\
$\lambda _{2}=k_{31}+k_{23}k_{12}/(k_{12}+k_{21})$~~~~~~~(2)\\
$a=k_{23}k_{12}/\left [ k_{31} (k_{12}+k_{21})\right ]$\\
\end{center}

Note that the usual  \textquotedblleft\textit{start-stop}\textquotedblright setup that serves to measure the $g^{(2)}$ function is only valid on timescale smaller than $R^{-1}$, where $R$ is the counting rate \cite{Reynaud1983}. In order to capture the full dynamics of the GQD photophysics, and especially the long bunching characteristics time ($\lambda{_{2}}^{-1}\sim\mu$s), we used the Hanbury Brown and Twiss (HBT) setup to measure $J(t)$ which is the histogram of photons detected at time $t$ provided that a photon is detected at time $t = 0$. To this end, a photon detection event on detector $D_{1}$ was used to trigger the acquisition of a PL time trace on detector $D_{2}$ using a wide-range time digitizer (P7887, FastComtec). After $N$ repetitions of the measurement, the resulting histogram $J(t)$ is directly linked to the $g^{(2)}(t)$ function through \cite{Huang2016, Martinez2016c}:
\begin{center}
$g^{(2)}(t) = \frac{J(t)}{NwR_{2}}$~~~~(3)
\end{center}
where $w$ is the time bin width and $R_{2}$ is the photon count rate on detector $D_{2}$.

Extended Data Fig.~\ref{EAFigphotophys}a represents a typical recorded $g^{(2)}(t)$ function over 50~$\mu$s with an excitation power of 10~$\mu$W. We observe that the $g^{(2)}(0)$ at zero time delay has a non-negligible residual value for excitation power larger than a few $\mu$W. This deviation from $g^{(2)}(0)=0$ is due to the shortening of the antibunching characteristic time ${\lambda_{1}}^{-1}$ with the incident power bringing it closer to the instrument response \cite{Wu2007b}. Using 6~ps supercontinuum laser pulses, the instrument response function (IRF) was independently measured as a 0.9~ns FWHM Gaussian function (Extended Data Fig.~\ref{EAFigphotophys}b). The measured $g^{(2)}$ function is then well fitted by the convolution of the IRF with Eq.~(1).

To gain insights into the transition rates, we measured the $g^{(2)}$ function of a GQD at different excitation powers. By fitting the $g^{(2)}$ functions, we obtain the values of $\lambda _{1},\lambda _{2}$ and $a$. We plot these values as a function of excitation power (Extended Data Fig~\ref{EAFigphotophys}c, d and e, respectively). From these plots, we can deduce all of the transition rates.

The population rate of excited state $k_{12}$ is linked to the excitation power $P$ by the relation:
\begin{center}
$k_{12} = \sigma P/h\nu$
\end{center}
where $\sigma$ is the absorption cross-section and $\nu$ is the excitation frequency. Combining this relation with Eq~.2, we fit the $\lambda _{1}$ values with a linear function (see Extended Data Fig.~\ref{EAFigphotophys}c). From this fit, we obtained the absorption cross-section $\sigma\simeq 1.0\times 10^{-14}$~cm$^{2}$. By extrapolating the linear fit to zero excitation power, we also obtain the relaxation rate of the excited state $k_{21}=0.28 \pm 0.02$~ns$^{-1}$. The lifetime of excited state $k{_{21}}^{-1}$ is thus calculated to be 3.55 $\pm$ 0.24~ns. This value is compatible with our time-resolved PL measurements leading to life times ranging from 3 to 6~ns depending on the GQD.

Using Eq.~3 and 4, the relaxation rate of metastable state $k_{31}$ is given by
\begin{center}
$k_{31}=\frac{\lambda _{2}}{1+a} = 0.053 \pm 0.001~\mu \text{s}^{-1}$
\end{center}

We then fit the $\lambda _{2}$ by combining Eq. 3 and 6. Treating $k_{23}$ as a constant, we obtain  $k_{23}=0.025 \pm 0.005$~$\mu$s$^{-1}$.

We now estimate the quantum yield of GQD. Using the transition rates, the detected fluorescence rate can be expressed as \cite{Wu2007b}
\begin{center}
$R=\eta _{det}\eta _{Q}\frac{k_{21}}{(k_{21}/k_{12}+k_{23}/k_{31}+1)}$
\end{center}
where $R$ is the sum of count rates on the two APDs, $\eta_\mathrm{det}$ is the overall detection efficiency and $\eta_Q$ is the fluorescence quantum yield. By fitting the fluorescence saturation curve, together with the deduced transition rates, we obtain $\eta _{det}\times \eta _{Q}\approx 0.05$. In Tab. 1  we list the transmission or collection efficiency of each optical component in the detection path, leading to an overall collection efficiency expected to be lower than $15\%$. Therefore, a lower bound of the fluorescence quantum yield is $\sim 35\%$.

\newcolumntype{Y}{>{\centering\arraybackslash}X}
\begin{table}[ht]
\centering
\begin{tabular}{|c|c|c|c|}
  \hline
  objective lens & mirrors & dichroic mirror & beamsplitter \tabularnewline

\hline

  30\% & 92\% & 98\% & 95\%\tabularnewline
\hline
lenses & optical filters & APD & $\eta_{det}$ \tabularnewline

\hline

97\% & 88\% & 70\% & $\sim$ 15\% \tabularnewline
\hline
\end{tabular}
\end{table}

\textbf{Data availability} The data that support the findings of this study are available from the corresponding author upon reasonable request.\\

********************%

\setcounter{figure}{0}
\renewcommand\figurename{\textbf{Extended Data Figure}}

\begin{figure*}
\includegraphics[scale=0.85]{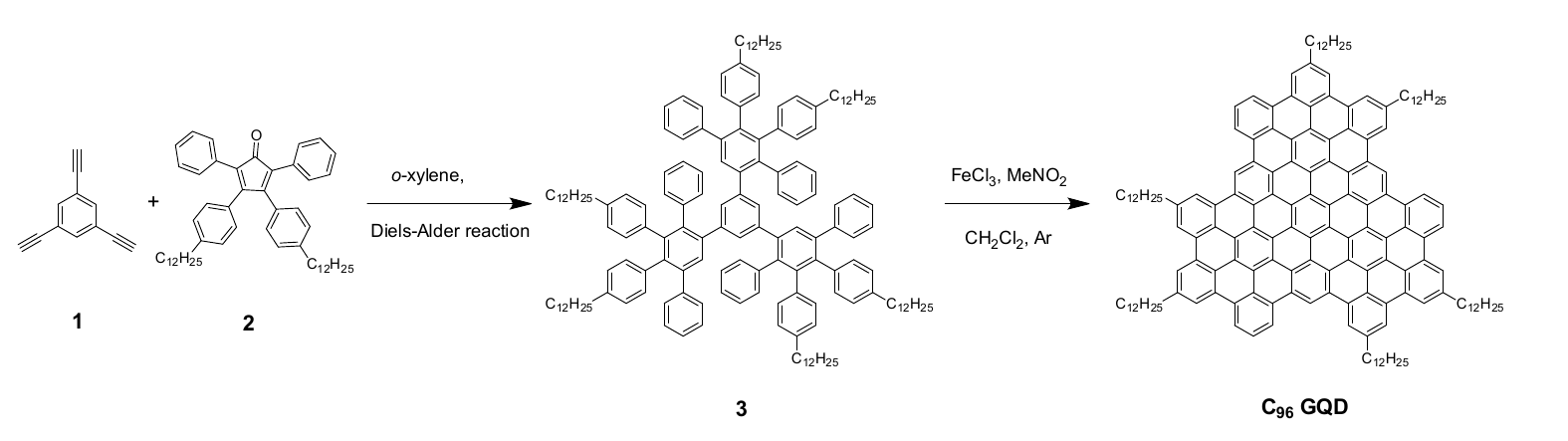}
\caption{Reaction route to the production of C$_{96}$ GQD.}
\label{EAFigChimie1}
\end{figure*}

\begin{figure*}
\includegraphics[scale=0.85]{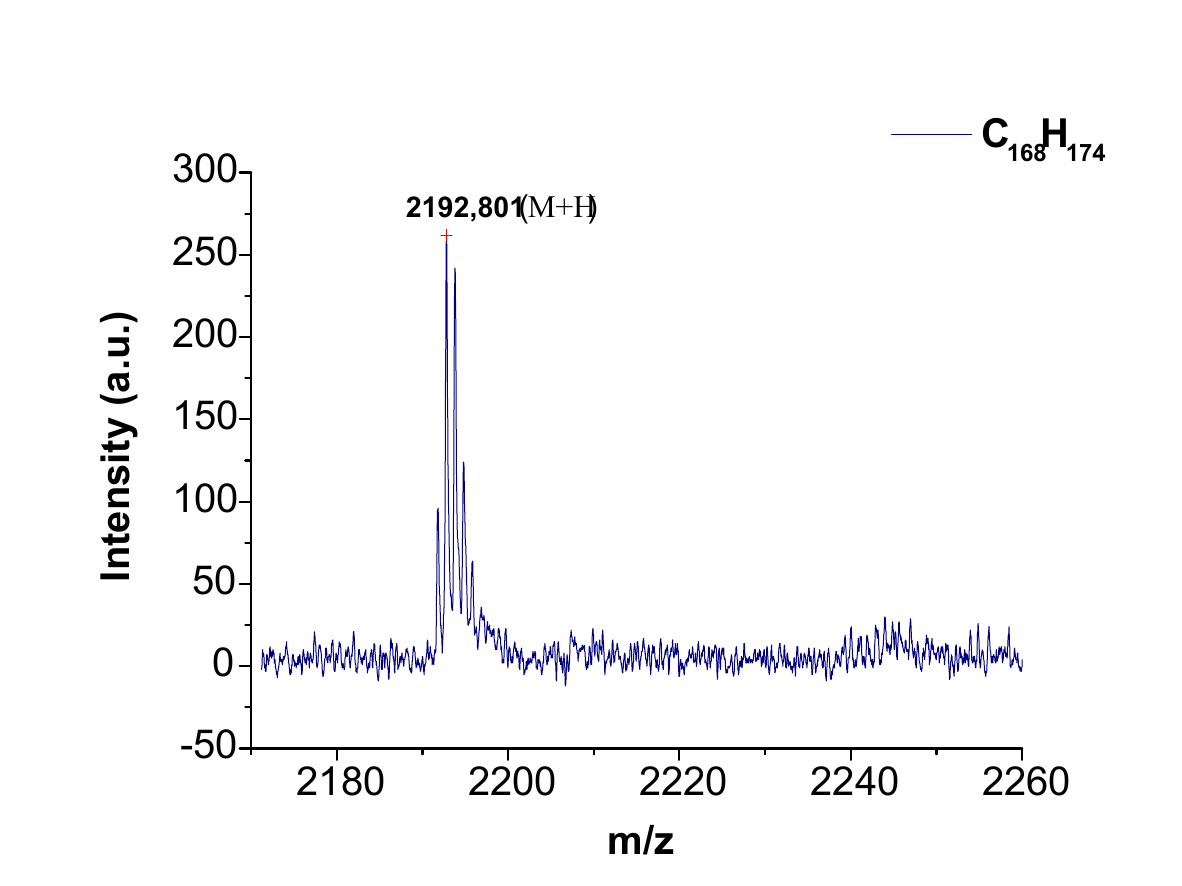}
\caption{\textbf{MALDI-TOF mass spectrum} of C$_{96}$ GQD. The spectrum shows the absence of starting or incompletely oxidized materials.}
\label{EAFigChimie2}
\end{figure*}

\begin{figure*}
\includegraphics[scale=0.55]{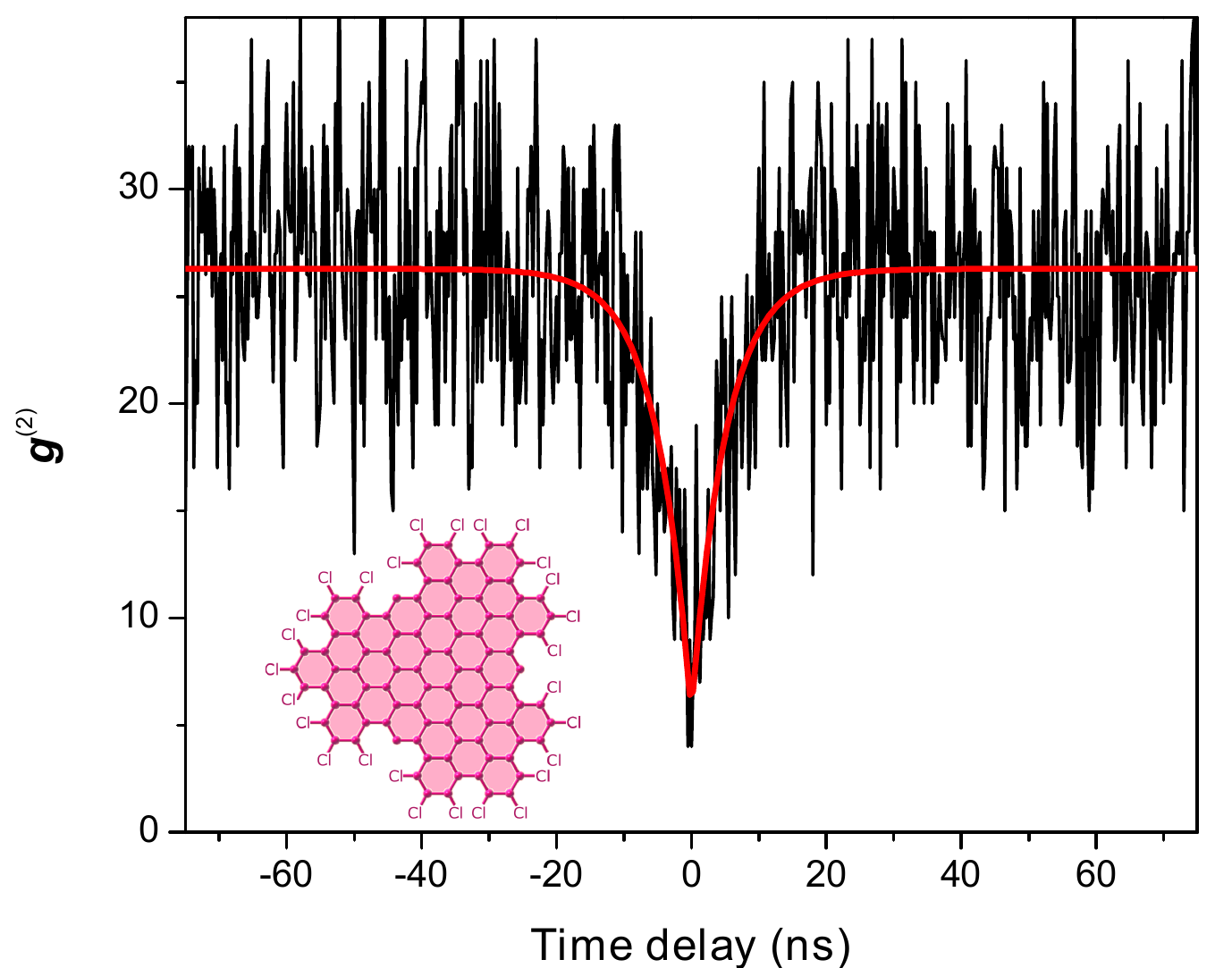}
\caption{\textbf{$g^{(2)}$ curve} of the C$_{96}$Cl GQD whose spectrum is displayed on Fig.1c. Here the edge chlorination leads to an almost 100~nm redshift of the main PL line while maintaining a single photon emission. Inset: Schematic chemical structure of the C$_{96}$Cl GQDs.}
\label{EAFigC96Cl}
\end{figure*}

\begin{figure*}
\includegraphics[scale=0.35]{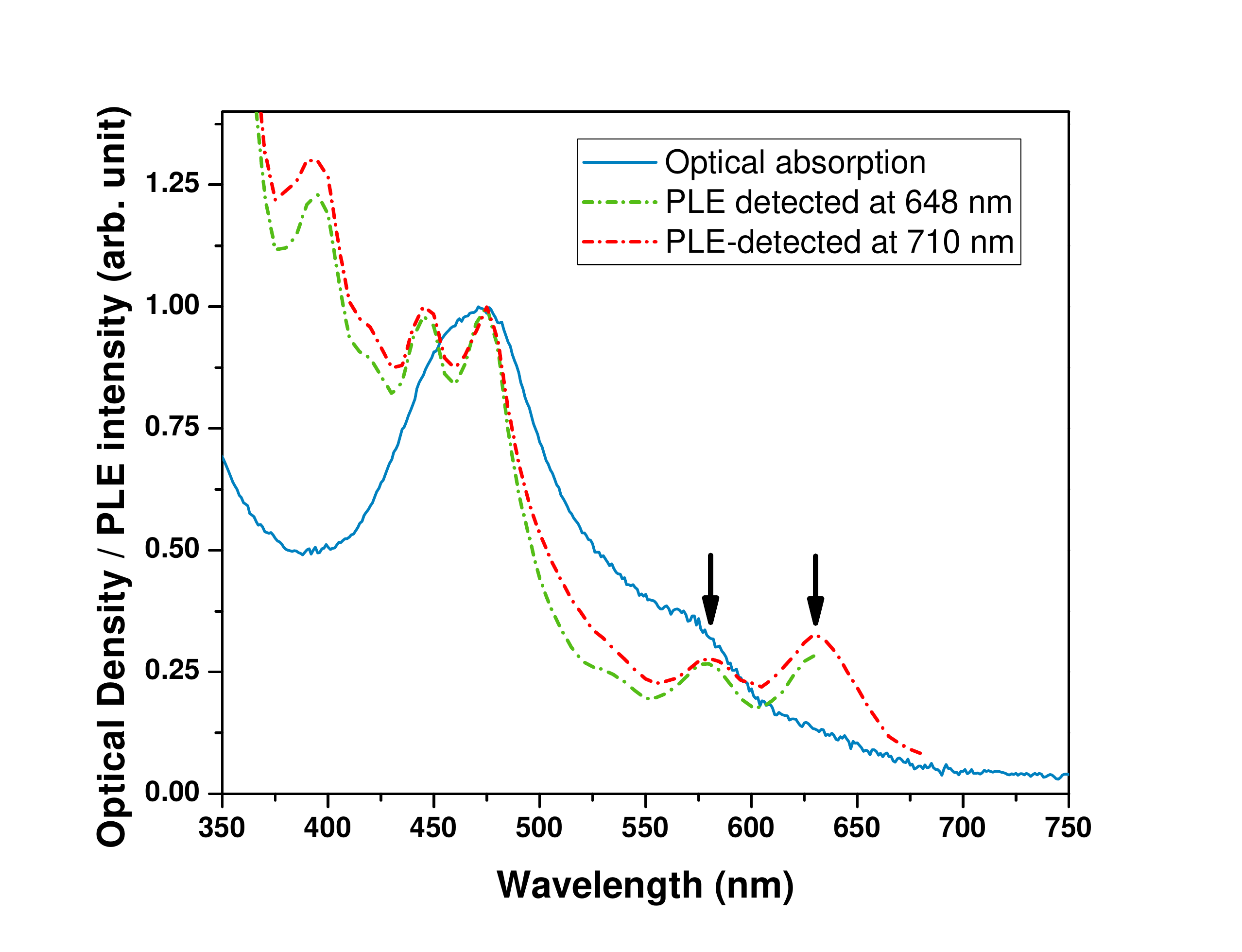}
\caption{\textbf{Optical absorption spectrum} of C$_{96}$ GQDs in solution in trichlorobenzene (blue) and PLE spectra (green and red). The absorption spectrum is composed of a main line at $\sim$475~nm and of some shoulder on the low energy tail. Likewise, both PLE spectra recorded on the two main PL lines follow the trend of the absorption spectrum. Moreover, PLE lines are much more structured than the absorption spectrum. This may be an indication that some aggregates still remain in the suspension. Their PL quantum yield would be much lower than the one of monomers, explaining why the PL spectrum is dominated by the monomers.}
\label{EAFig1}
\end{figure*}

\begin{figure*}
\includegraphics[scale=0.55]{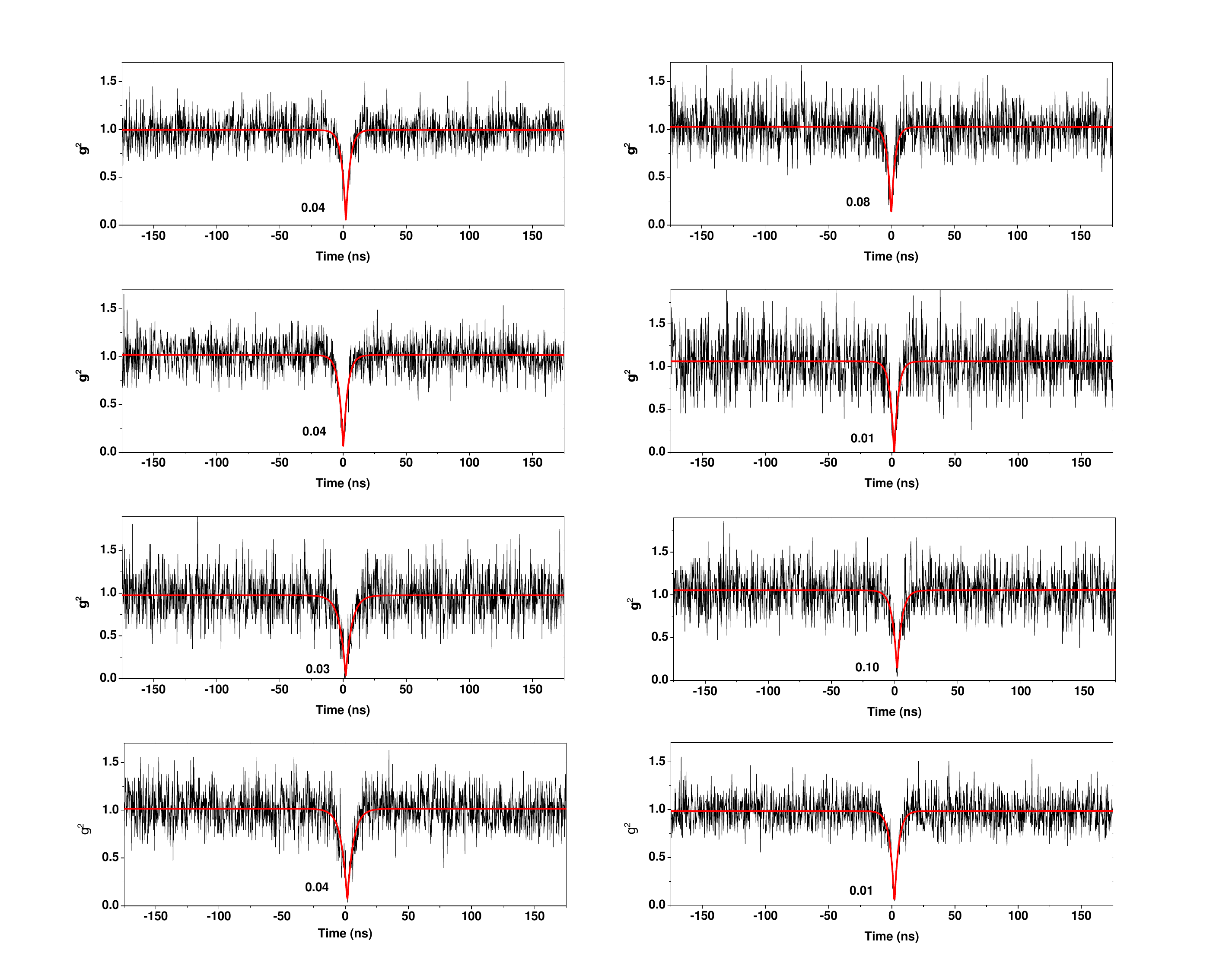}
\caption{\textbf{Collection of g$^2(\tau)$ traces} on different GQDs recorded at moderate excitation power ($\approx 200$~nW).}
\label{EAFig6}
\end{figure*}

\begin{figure*}
\includegraphics[scale=0.35]{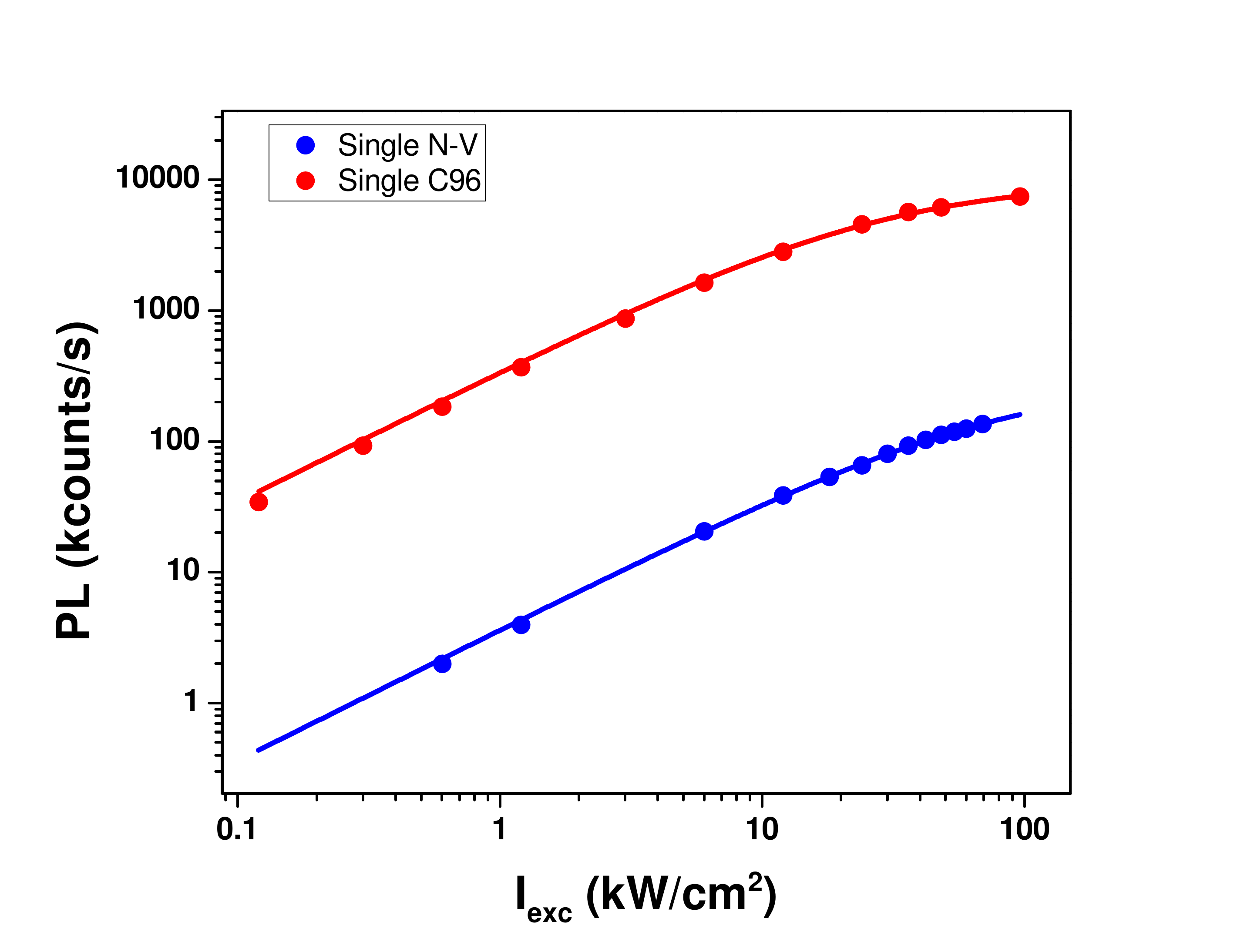}
\caption{\textbf{Saturation curves} of a GQD (red dots) and a single NV center in diamond (blue dots). Both were fitted with the saturation functions described in the main text.}
\label{EAFig2}
\end{figure*}

\begin{figure*}
\includegraphics[scale=1.5]{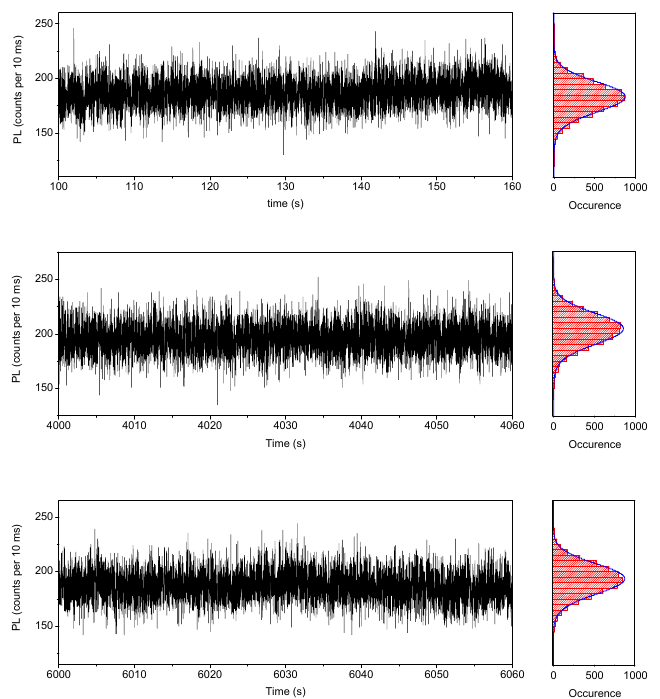}
\caption{\textbf{Intensity histograms} PL time traces (the same as shown in the zooms of Fig. 2d of main text) and corresponding count rate histograms. The histograms were fitted by normal functions (blue curves)}
\label{EAFig3}
\end{figure*}

\begin{figure*}
\includegraphics[scale=.1]{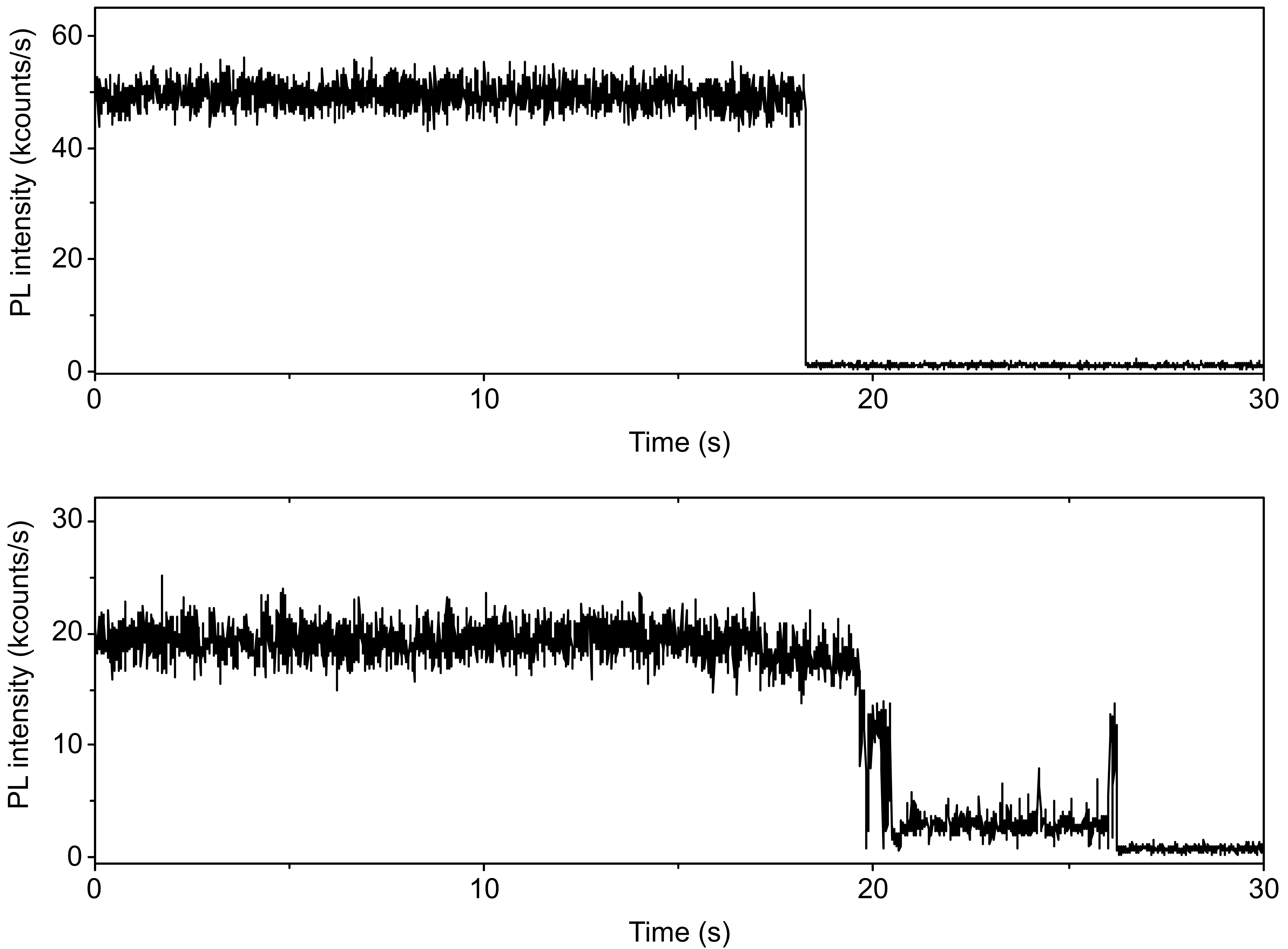}
\caption{\textbf{PL time traces} recorded from different GQDs around the time of bleaching. Bin size: 10 ms. }
\label{EAFig4}
\end{figure*}

\begin{figure*}
\includegraphics[scale=0.55]{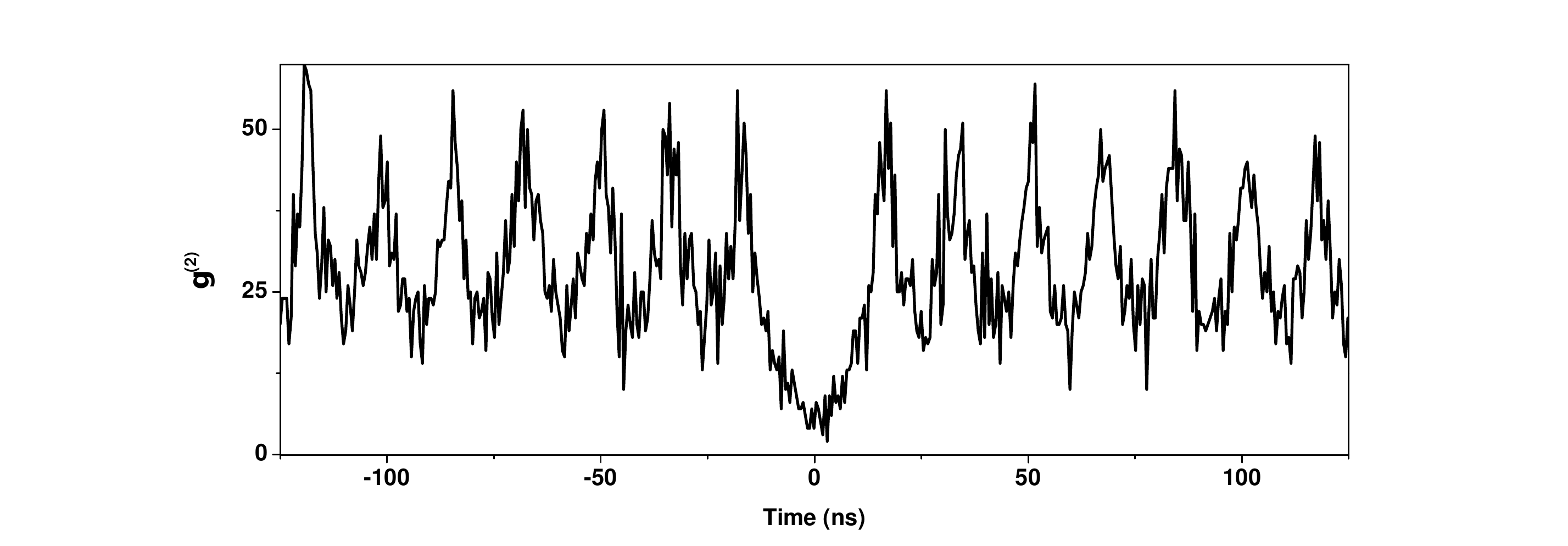}
\caption{\textbf{g$^2(\tau)$ trace under pulsed excitation} (580 nm) on the same GQD as the one corresponding to the PL decay curve of Fig. 2b of the main text.}
\label{EAFig5}
\end{figure*}

\begin{figure*}
\includegraphics[scale=0.225]{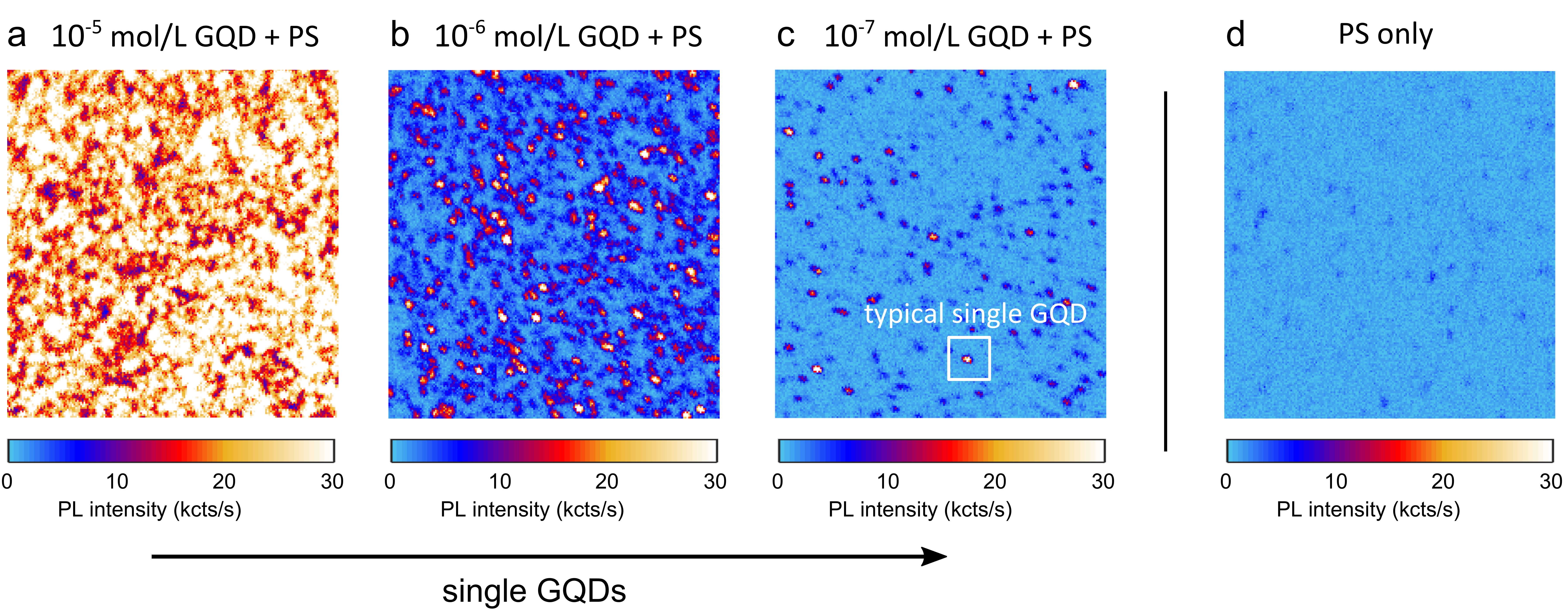}
\caption{\textbf{PL maps of GQDs} in a polystyrene matrix spin-coated from solutions at different concentrations of C$_{96}$ GQD: \textbf{a} 10$^{-5}$mol.L$^{-1}$; \textbf{b} 10$^{-6}$mol.L$^{-1}$; \textbf{c} 10$^{-7}$mol.L$^{-1}$. \textbf{d} PL map of polystyrene film without GQDs. Map size: $20\times 20$~$\mu$m$^2$. When a highly concentrated solution is used to make the thin film, luminescence is observed everywhere on the sample. When the concentration decreases, the spatial distribution of PL is more and more localized. It ensures that the bright luminescence spots do arise from GQDs. Moreover, the \textbf{d} panel shows a map of a polystyrene film without GQDs. It shows very weak PL signal in comparison with the GQD samples. The localized weak PL spots that are still observed bleach quasi instantaneously, in strong contrast with the GQDs spots that are stable for hours.}
\label{EAFigPLmaps}

\end{figure*}

\begin{figure*}
\includegraphics[scale=0.175]{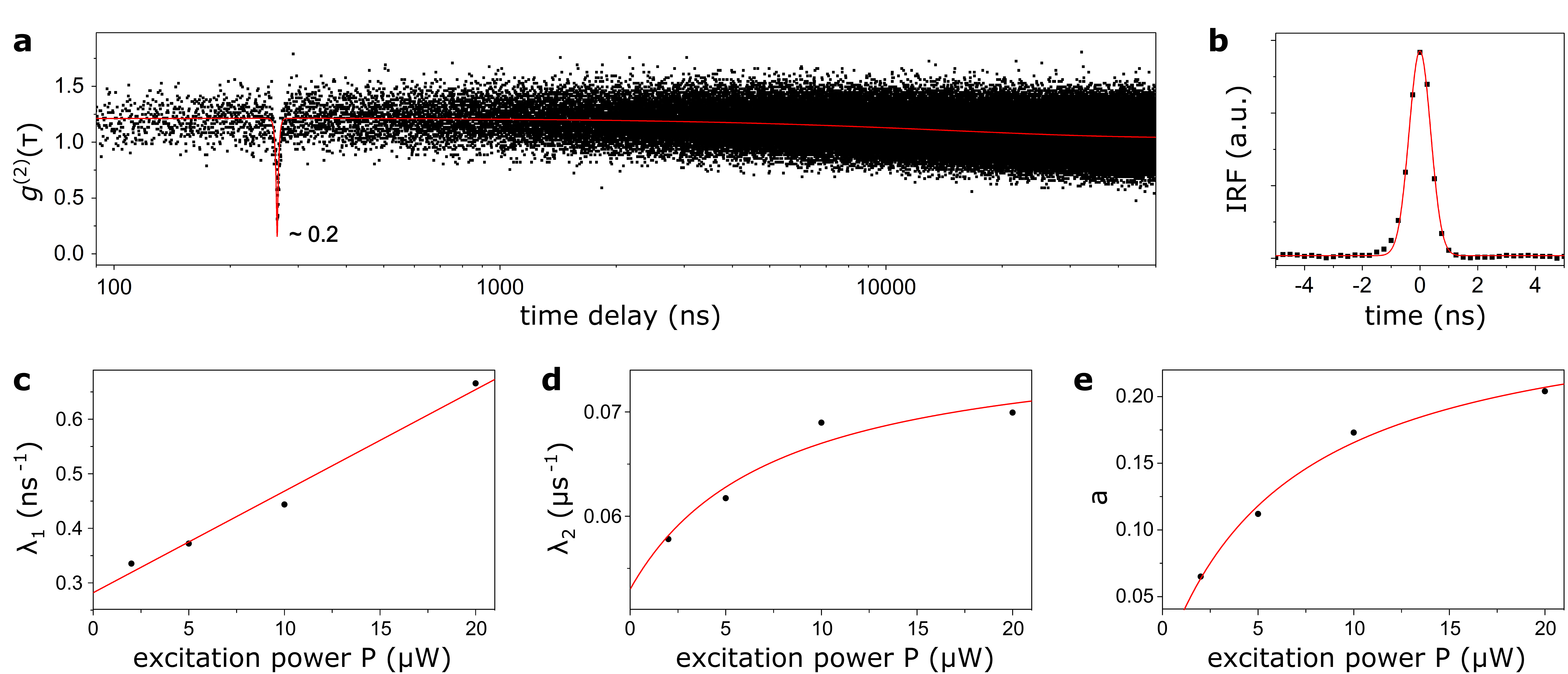}
\caption{\textbf{Photophysics of a single GQD}. \textbf{a,} Long-time-scale g$^{(2)}$ function recorded with an excitation power of 10~$\mu$W. Bin width: 250~ps. A time delay is introduced for log-scale display. The red line is a fit by the convolution of Eq.1 and the measured instrument response function (IRF). \textbf{b,} Impulse Response Function of the detector measured using 6-ps laser pulses. The Gaussian fit (red line) gives a FWHM of 0.9~ns. \textbf{c-e,} Intensity dependence of the fit parameters $\lambda_{1}$ (c), $\lambda_{2}$ (d) and $a$ (e) of $g^{(2)}$. The red lines are fits according to the three-level model.}
\label{EAFigphotophys}

\end{figure*}

\end{document}